\documentclass[12pt]{article}

\begin{document}

\author{A.I.Volokitin$^{1,2}$\footnote{Corresponding author. 
\textit{E-mail address}:avoli@samgtu.ru}    and B.N.J.Persson$^1$ \\
\\
$^1$Institut f\"ur Festk\"orperforschung, Forschungszentrum \\
J\"ulich, D-52425, Germany\\
$^2$Samara State Technical University, 443100 Samara,\\
Russia}
\title{Quantum field theory of the van der Waals friction}
\maketitle

\begin{abstract}
The van der Waals friction between two semi-infinite solids, a small neutral 
particle and semi-infinite solid is reconsidered on the basis of thermal quantum 
field theory in Matsubara formulation. The friction calculated in this approach is 
in agreement with the friction calculated in the framework of dynamical modification 
of the Lifshitz theory with use of the fluctuation-dissipation theorem. This 
solves the problem about the applicability of the Lifshitz theory to the dynamical 
situation. In quantum field theory the calculation of the friction to linear 
order in the sliding velocity is reduced to the finding of the equilibrium 
Green functions. Thus this approach can be extended for bodies with complex geometry. 
We show that the van der Waals friction can be measured in non-contact friction 
experiment using state-of-the-art equipment.    
\end{abstract}

\section{ Introduction}

 A great deal of
attention has been devoted to the problem of non-contact friction  between 
nanostructures, including, for example, the frictional drag
force between electrons in two-dimensional quantum wells 
\cite{Gramila1,Gramila2,Sivan},
and the friction force between an atomic force microscope tip and a
substrate \cite{Dorofeev,Gostmann,Stipe,Mamin,Hoffmann,Volokitin4,Volokitin5,
Volokitin8,Volokitin9,Kuehn}. This is because of
it importance for ultrasensitive force detection experiments. 
The ability to detect small forces is inextricably linked to
friction via the fluctuation-dissipation theorem. According to this theorem,
the random force that makes a small particle jitter would also cause
friction if the particle was dragged through the medium. For example, the
detection of single spins by magnetic resonance force microscopy \cite{Rugar}
, which has been proposed for three-dimensional atomic imaging \cite{Sidles}
and quantum computation \cite{Berman}, will require force fluctuations (and
consequently the friction) to be reduced to unprecedented levels. In
addition, the search for quantum gravitation effects at short length scale
\cite{Arkani}, and future measurements of the  Casimir and van der Waals 
forces \cite
{Mohideen}, may eventually be limited by non-contact friction effects.

In non-contact friction the bodies are separated by a potential barrier
thick enough to prevent electrons or other particles with a finite rest mass
from tunneling across it, but allowing interaction via the long-range
electromagnetic field, which is always present in the gap between bodies. 
The presence of an inhomogeneous tip-sample
electric fields is difficult to avoid, even under the best experimental
conditions \cite{Stipe}. For example, even if both the tip and the sample
were metallic single crystals, the tip would still have corners, and more
than one crystallographic plane exposed. The presence of atomic steps,
adsorbates, and other defects will also contribute to the spatial variation
of the surface potential. This is referred to as ``patch effect''. The
surface potential can also be easily changed by applying a voltage between
the tip and the sample. An inhomogeneous electric field can also be created
by charged defects embedded in a dielectric sample. The relative motion of
the charged bodies will produce friction which will be denoted as the
\textit{electrostatic friction}.

The electromagnetic field can also be created by the fluctuating current
density due to thermal and quantum fluctuations inside the bodies. This
fluctuating electromagnetic field gives rise to the well-known long-range
attractive van der Waals interaction between two bodies \cite{Lifshitz2,Lifshitz1},
and is responsible for radiative heat transfer. If the bodies
are in relative motion, the same fluctuating electromagnetic field will give
rise to a friction which will be denoted as the \textit{van der Waals friction}.

The origin of the van der Waals friction is closely connected with the van
der Waals interaction. The van der Waals interaction arises when an atom or
molecule spontaneously develops an electric dipole moment due to  quantum
fluctuations. The short-lived atomic polarity can induce a dipole moment in a
neighboring atom or molecule some distance away. The same is true for
extended media, where thermal and quantum fluctuation of the current
density in one body induces a current density in other body; the interaction
between these current densities is the origin of the van der Waals
interaction. When two bodies are in relative motion, the induced current
will lag slightly behind the fluctuating current inducing it, and this is
the origin of the van der Waals friction. 

The van der Waals interaction is
mostly determined by exchange by virtual photons between the bodies
(connected with quantum fluctuations), and does not vanish even at zero
temperature. On the other hand, the van der Waals friction, at least to lowest
order of perturbation theory, and to linear order in the sliding velocity,
is determined by exchange of real photons, and vanishes at zero temperature.

To clarify the origin of the van der Waals friction let us consider two flat
parallel surfaces, separated by a sufficiently wide vacuum gap, which
prevents electrons from tunneling across it. If the surfaces are in relative
motion (velocity $v$) a frictional stress will act between them. This
frictional stress is related with an asymmetry of the reflection coefficient
along the direction of motion; see Fig.1. If one body emits radiation, then
in the rest reference frame of the second body these waves are Doppler
shifted which will result in different reflection coefficients. The same is
true for radiation emitted by the second body. The exchange of ``Doppler
shifted photons'' is the origin of van der Waals friction.

From the point of view of the quantum mechanics the van der Waals friction
originates from two types of processes:(a) Photons are created in each body
with opposite momentum and the frequencies of these photons are connected by
$vq_x=\omega _1+\omega _2$, where $q_x$ is  the momentum transfer. (b) A photon is
annihilated in one body and created in another. The first process (a) is
possible even at zero temperature, and it gives rise to a friction force
which depends cubically on sliding velocity \cite{Pendry2,Volokitin6}. The
second process (b) is possible only at finite temperatures, and gives rise
to a friction which depends linearly on the sliding velocity. Thus, process
(b) will give the main contribution to the friction at sufficiently high
temperatures, and at not too large velocities.

In contrast to the van der Waals interaction, for which theory is well
established, the field of van der Waals friction is still controversial. As
an example, different authors have studied the van der Waals friction
between two flat surfaces in parallel relative motion using different
methods, and obtained results which are in sharp contradiction to each
other. The first calculation of van der Waals friction was done by
Teodorovich \cite{Teodorovich}. Teodorovich assumed that the force of
friction can be calculated as the ordinary van der Waals force between
bodies at rest, whose dielectric function depend on the velocity due to the
Doppler shift. However, from the dynamical modification of the Lifshitz's
theory it follows \cite{Volokitin6} that it is not true. Later the same
approach was used by Mahanty \cite{Mahanty} to calculate the friction
between molecules. Both theories predict wrong nonzero friction at absolute
zero of temperature to linear order in the sliding velocity. The same nonzero 
linear friction at zero temperature was
predicted in \cite{Dedkov1,Dedkov2}. In \cite{Volokitin7} it was shown that
the basic equation in \cite{Dedkov1,Dedkov2} is incorrect, and a correct
treatment gives a vanishing linear friction at $T=0$K. Schaich and Harris
developed a theory \cite{Schaich1} which describes the dependence of
friction on the temperature $T$ and on the separation $d$. However in their
calculations they made the series of unphysical approximations, and 
 for the Drude model their final formula for the friction for parallel
relative motion gives a divergent result. The friction obtained in \cite
{Levitov, Polevoi,Mkrtchian} vanishes in the limit of infinite light
velocity $c\rightarrow \infty $. However, at least for short distances, one
can neglect by retardation effects when calculating the van der Waals
friction, as well as van der Waals interaction. Pendry \cite{Pendry2}
assumed zero temperature  and  neglected retardation
effects, in which case the
friction depends cubically on the velocity. Persson and Zhang \cite{Persson 
and Zhang} obtained the formula for
friction in the limit of small velocities and finite temperature using a simple 
quantum mechanical approach,  again
neglecting  retardation effects. In Ref.\cite{Volokitin7} Volokitin and
Persson developed a theory of the van der Waals friction based on the
dynamical modification of the well known Lifshitz theory \cite{Lifshitz2} of
van der Waals interaction. In the nonretarded limit and for zero temperature
this theory agrees with the results of Pendry \cite{Pendry2}. Similarly, in
the nonretarded limit and for small sliding velocity this theory agrees with
the study of Persson and Zhang \cite{Persson and Zhang}. In \cite
{Volokitin4, Volokitin5} the theory was extended to two flat surfaces in
normal relative motion. For the case of resonant photon tunneling between
surface localized states, normal motion results in drastic difference from
parallel relative motion.  It was shown that the friction may increase
by many orders of magnitude when the surfaces are covered by adsorbates, or
can support low-frequency surface plasmons. In this case the friction is
determined by resonant photon tunneling between adsorbate vibrational modes,
or surface plasmon modes. When one of the bodies is sufficiently rarefied,
this theory gives the friction between a flat surface and a small particle,
which in the nonretarded limit agrees with the results of Tomassone and
Widom \cite{Tomassone}. A theory of the van der Waals friction between a
small particle and flat surface, which takes into account screening,
nonlocal optic effects, and retardation effects, was developed in \cite
{Volokitin7}.

 Dorofeyev \textit{et.al.} \cite{Dorofeev} claim that the
non-contact friction observed in \cite{Dorofeev,Gostmann} is due to Ohmic
losses mediated by the fluctuating electromagnetic field. This claim is
controversial, however, since the van der Waals friction for good conductors
like copper has been shown \cite{Volokitin6,Volokitin7,Persson
and Volokitin} to be many orders of magnitude smaller than the friction
observed by Dorofeyev \textit{et.al.}. In \cite{Greffet2} it was proposed
that the non-contact friction observed in Ref.\cite{Stipe} can be explained by 
the van der Waals friction between the high
resistivity mica substrate and silica tip. However in the experiment 
the mica substrate and
silica tip were coated by gold films thick enough to completely screen the
electrodynamic interaction between the underlying dielectrics.

At  small separation $d\sim 1$nm, resonant photon tunneling between
adsorbate vibrational modes on the tip and the sample may increase the
van der Waals friction by seven orders of magnitude in comparison with 
the good conductors with clean 
surfaces \cite{Volokitin4, Volokitin5}. However, the distance dependence ($
\sim 1/d^6$) is stronger than observed experimentally \cite{Stipe}.

Recently, a theory of non-contact friction was suggested where the friction
arises from Ohmic losses associated with the electromagnetic field created
by moving charges induced on the surface of the atomic force microscope 
tip by the bias voltage  or spatial variation of the surface potential 
\cite{Volokitin8,Volokitin9}.
It was  shown that the electrostatic friction can
be greatly enhanced if there is an incommensurate adsorbate layer which 
can exhibit
acoustic vibrations. This theory gives a tentative explanation for the
 experimental non-contact friction data \cite{Stipe}.

 Although at present there are many theories of the van der Waals
friction, which are frequently contradict to each other, the rigorous theory
based on the quantum field theory is still not developed. For the van der Waals
interaction such quantum field theory was developed in Ref.\cite{Lifshitz1}. On the
base of this theory were solved some problems which can not be solved in the
Lifshitz theory of the van der Waals interaction. Due to of great importance of the 
van der Waals friction for the understanding of the origin of the non-contact friction 
 there is strong demand for the creation of the rigorous theory of the 
van der Waals friction.

This article is organized as follows. In Section 2 we present a
short overview of the basic idea of the quantum field theory
of the van der Waals friction. We apply this theory for the calculation 
of the van der Waals friction between two semi-infinite solids (Sec.3), 
and a small particle and semi-infinite solid (Sec.4), for parallel and 
normal relative motion. These calculations confirm early results obtained 
with use of the dynamical modification of the Lifshitz theory 
and the fluctuation-dissipation theorem. 
Thus these calculations solves the problem of applicability of the Lifshitz theory 
for bodies in relative motion. The quantum field theory of the van der Waals 
friction is more general and can be applied for bodies with complex geometry. 
In Sec.5 we show that
the van der Waals friction can be greatly enhanced for high-resistivity 
metals, dielectrics with strong absorption in low-frequency region, and for two-
dimensional systems, e.g.
2D-electron systems on the dielectric substrate, 
or  incommensurate layers of adsorbed ions exhibiting
acoustic vibrations. The origin of this enhancement is related to the fact
that screening in  2D-systems is much less effective than for
3D-systems. The fluctuating charges  will induce ``image''
charges in the 2D-system. Because of the
finite response time during relative motion of the bodies this ``image''
charge will lag behind the charge which induce it, and this
 results in a force friction acting on the bodies.  The weaker screening  in
2D-systems  will results in  larger lag of the ``image'' charge in  2D-systems as
as compare  3D-systems, and, as the consequence, in larger friction.
 Sec.6 presents the conclusions and the
outlook. Appendix A contains the details of the derivations of 
the Green functions of the electromagnetic field in two plane surface 
geometry. These Green functions are used in Secs. 3 and 4.

\section{General formalism}

There are two approaches to the theories of the van der Waals interaction
and the van der Waals friction. In the first approach the fluctuating
electromagnetic field is considered as a classical field which can be
calculated from Maxwell's equation with the fluctuating current density as
the source of the field, and with appropriate boundary conditions. This
approach was used by Lifshitz in the theory of the van der Waals interaction
\cite{Lifshitz2} and by Volokitin and Persson for the  van der Waals
friction \cite{Volokitin5,Volokitin6}. The calculation of the van der Waals
friction is more complicated than of the van der Waals force because it requires the
determination of the electromagnetic field between  moving boundaries. The
solution can be found by writing the boundary conditions on the surface of
each body in the rest reference frame of this body. The relation between the
electromagnetic fields in the different reference frames is determined by
the Lorenz transformation. The advantage of this approach is in that, in
principle, it can be used for the calculation of friction at the arbitrary
relative velocities. However, the calculations become very complicated for
bodies with complex geometry. At present the solutions are known for the van
der Waals friction between two parallel plane surfaces \cite
{Volokitin5,Volokitin6}, and between a small particle and plane surface \cite
{Volokitin7}.

In the second approach the electromagnetic field is treated in the frame of
the quantum field theory \cite{Abrikosov}. This approach was used in Ref.\cite{Lifshitz1} to
obtain the van der Waals interaction for an arbitrary inhomogeneous medium
all parts of which are at rest.

For two bodies in slow uniform relative motion (velocity $\mathbf{v}$) the
force acting on either body may be written as $\mathbf{F}=\mathbf{F}_0-%
\stackrel{\leftrightarrow }{\mathbf{\Gamma }}\cdot \mathbf{v}$, where the
adiabatic force $\mathbf{F}_0$ is independent of $\mathbf{v}$, and $%
\stackrel{\leftrightarrow }{\mathbf{\Gamma }}$, the so-called friction
tensor, is  defined by
\begin{equation}
\stackrel{\leftrightarrow }{\mathbf{\Gamma }}=(k_BT)^{-1}\mathrm{Re}%
\int_0^\infty dt\left\langle \hat {\mathbf{F}}(t)\hat {\mathbf{F}%
}(0)\right\rangle  \label{Kubo}
\end{equation}
Here $\left\langle ...\right\rangle $ represents a thermal average of the
fluctuating force in the equilibrium state at fixed separation $d$ between
bodies, and $\hat {\mathbf{F}}(t)$ is the force operator in the Heisenberg
representation.  Eq. (\ref{Kubo}) is a consequence of the
fluctuation-dissipation theorem \cite{Landau}. For  the interaction
between a localized and an extended system, Eq.(\ref{Kubo}) has been derived
by several authors (Schaich 1974 \cite{Schaich}, d'Agliano \textit{et al}
1975 \cite{d'Agliano}, Nourtier 1977 \cite{Nourtier}) and is also valid for
two extended systems. In the context of the van der Waals friction Eq.(\ref{Kubo})
 was  used by Schaich and Harris \cite{Schaich1}, but their treatment
is incomplete.

In the case of  extended systems the fluctuating force operator can be
expressed through the operator of the stress tensor $\hat \sigma _{ik}$
\begin{equation}
\hat F_i=\int dS_k\hat \sigma _{ik},  \label{stress1}
\end{equation}
where the integration is over the surface of the one of the bodies and
\begin{equation}
\hat \sigma _{ik}=\frac 1{4\pi }\left[ E_iE_k+B_iB_k-\frac 12\delta
_{ik}\left( E^2+B^2\right) \right]  \label{stress2}
\end{equation}
where $E_i$ and $B_i$ are the electric and magnetic induction field
operator, respectively. The calculation of the force-force correlation
function can be done using the methods of the quantum field theory \cite
{Abrikosov,Mahan}. Such calculations are described in Sec. 3 for two
plane parallel surfaces, and in Sec.4 for a small particle and plane 
surface, for parallel and normal relative motion. The
advantage of this approach is  that it only involves   finding of the
Green's functions of the electromagnetic field for the equilibrium system
with fixed boundaries. Thus, this approach can be easily extended to bodies
with complex geometry. However, it is restricted to small relative
velocities.

\section{Van der Waals friction between two plane surface}

\subsection{Parallel relative motion}

Assume that the $xy$-plane coincides with one of the surface. For
parallel relative motion the friction coefficient $\Gamma _{\Vert }=\Gamma
_{xx}=\Gamma _{yy}$. Using the methods of  quantum field theory \cite
{Abrikosov} the expression for the friction coefficient (\ref{Kubo}) for
parallel relative motion can be written in the form
\begin{equation}
\Gamma _{\Vert }=\lim_{\omega _0\to 0}\mathrm{Im}\frac{G_{xx}^R(\omega
_0+i\delta )}{\omega _0},
\label{d1}
\end{equation}
where $G_{xx}^R$ is the retarded Green's function determined by
\begin{equation}
G_{xx}^R(\omega )=\frac i\hbar \int_0^\infty dte^{i\omega t}\left\langle
\hat F_x(t)\hat F_x(0)-\hat F_x(0)\hat F_x(t)\right\rangle
\end{equation}
where
\begin{equation}
\hat F_x=\int dS_z\hat \sigma _{xz},
\end{equation}
where the surface integral is taken over the surface of the body at $z=0$,
\begin{equation}
\hat \sigma _{xz}=\left(E_xE_z+E_zE_x+B_xB_z+B_zB_x\right)/8\pi 
\end{equation}
The function $G_{xx}^R$ can be obtained by analytic continuation in the
upper half plane of the temperature Green's function $G_{xx}$ determined on
the discrete set of point $i\omega _n=i2\pi n/\beta $ by the formula
\begin{equation}
G_{xx}(i\omega _n)=-\frac 1\hbar \int_0^\beta d\tau e^{i\omega _n\tau
}\left\langle T_\tau \hat F_x(\tau )\hat F_x(0)\right\rangle ,
\end{equation}
where $n$ is an integer  and $\beta =\hbar /k_BT$. $T_\tau $ is the time-
ordering operator. The function $G_{xx}(i\omega_n)$ can be calculated using standard 
techniques
of  quantum field theory \cite{Abrikosov,Mahan}. Thus, the
function $G_{xx}(i\omega_n)$ can be represented through the Green's functions of the
electromagnetic field
\begin{equation}
 D_{ij}^{EE}(\mathbf{r},\mathbf{r}^{\prime },i\omega_n) = 
D_{ij}(\mathbf{r},\mathbf{r}^{\prime },i\omega_n) =
-\frac 1\hbar \int_0^\beta d\tau e^{i\omega _n\tau
}\left\langle T_\tau \hat E_i(\tau )\hat E_j(0)\right\rangle ,
\end{equation}
where the retarded Green functions 
$D_{ij}(\mathbf{r},\mathbf{r}^{\prime },\omega)$ obey the equations 
\cite{Abrikosov}
\begin{eqnarray}
&&\left( \nabla _i\nabla _k-\delta _{ik}\nabla ^2\right) D_{kj}(\mathbf{r},
\mathbf{r}^{\prime },\omega )-(\omega /c)^2\int d^3x^{\prime \prime
}\epsilon _{ik}(\mathbf{r},\mathbf{r}^{\prime \prime },\omega )D_{kj}(
\mathbf{r}^{\prime \prime },\mathbf{r}^{\prime },\omega)  \nonumber \\
&=&(4\pi \omega ^2/c^2)\delta _{ij}\delta (\mathbf{r}-\mathbf{r}^{\prime })
\label{green1}
\end{eqnarray}
\begin{eqnarray}
&&\left( \nabla _j^{\prime }\nabla _k^{\prime }-\delta _{jk}\nabla ^{\prime
2}\right) D_{ik}(\mathbf{r},\mathbf{r}^{\prime },\omega)-(\omega /c)^2\int
d^3x^{\prime \prime }\epsilon _{kj}(\mathbf{r}^{\prime \prime },\mathbf{r}
^{\prime },\omega)D_{ik}(\mathbf{r},\mathbf{r}^{\prime \prime },\omega)
\nonumber \\
&=&(4\pi \omega ^2/c^2)\delta _{ij}\delta (\mathbf{r}-\mathbf{r}^{\prime })
\label{green2}
\end{eqnarray}
For the plane surface it is convenient to decompose the electromagnetic
field into  $s$- and $p-$ polarized plane waves. In
this representation with
$\hat q=\mathbf{q}/q$ and $\hat n=[\hat z\times \hat q]$,
where $\mathbf{q}$ is the surface component of the wave vector,
the Green's tensor is given by
\[
\stackrel{\leftrightarrow }{\mathbf{D}}^{EE}(\mathbf{r},\mathbf{r}^{{\prime }
})=\int \frac{d^2q}{(2\pi )^2}\Big(\hat nD_{nn}^{EE}(z,z^{\prime },\mathbf{q}
)\hat n+\hat qD_{qq}^{EE}(z,z^{\prime },\mathbf{q})\hat q+\hat
zD_{zz}^{EE}(z,z^{\prime },\mathbf{q})\hat z
\]
\begin{equation}
+\hat zD_{zq}(z,z^{\prime },\mathbf{q})\hat q
+\hat qD_{qz}(z,z^{\prime },\mathbf{q})\hat z\Big)e^{i\mathbf{q\cdot }(
\mathbf{x}-\mathbf{x}^{\prime })}  \label{athree}
\end{equation}
where we have taken into account that $D_{nz}^{EE}=D_{nq}^{EE}=0$ (see Appendix A). 
For two plane surface geometry the solution of Eqs.(\ref{green1}, \ref{green2}) 
is described in Appendix A.
Using the methods of the quantum field theory \cite{Abrikosov,Mahan}
 for the Green function $G_{xx}$ we get
\[
G_{xx}(i\omega _n)=\frac{\hbar A}{16\pi ^2\beta }\int \frac{d^2\mathbf{q}}{
(2\pi )^2}\sum_{\omega _m}\frac{q_x^2}{q^2}\Big[D_{qq}^{EE}(\mathbf{q}
,i\omega _m,z,z^{\prime })D_{zz}^{EE}(-\mathbf{q},i\omega _n-i\omega
_m,z,z^{\prime })
\]
\[
+D_{qz}^{EE}(\mathbf{q},i\omega _m,z,z^{\prime })D_{zq}^{EE}(-\mathbf{q}
,i\omega _n-i\omega _m,z,z^{\prime })
\]
\[
+D_{qq}^{BB}(\mathbf{q},i\omega _m,z,z^{\prime })D_{zz}^{BB}(-\mathbf{q}%
,i\omega _n-i\omega _m,z,z^{\prime })
\]
\begin{equation}
+D_{qz}^{BB}(\mathbf{q},i\omega _m,z,z^{\prime })D_{zq}^{BB}(-\mathbf{q}%
,i\omega _n-i\omega _m,z,z^{\prime })\Bigg]_{z=z^{\prime }=0}  \label{C1one}
\end{equation}
where $A$ is the surface area, and   
$D_{ij}^{BB}(\mathbf{q},i\omega _m,z,z^{\prime })$ is given by
  \cite
{Abrikosov}
\begin{equation}
D_{ij}^{BB}(\mathbf{r},\mathbf{r}^{\prime },i\omega _n)=-\left( \frac
c{\omega _n}\right) ^2\mathrm{rot} _{ik}\mathrm{rot} _{jl}^{\prime }D_{kl}^{EE}
(\mathbf{r},
\mathbf{r}^{\prime },i\omega _n)
\end{equation}
In Eq.(\ref{C1one}) we omitted  terms involving product  of  Green's
functions associated with the $p$- and $s$- polarized electromagnetic field
because after summation they cancel to each other. Even without any 
detailed 
calculations it is clear that such terms must give zero contribution to the
friction because the $p$- and $s$- polarized waves must give independent
contributions to the friction.

All the sums over $\omega_m$ in Eq.(\ref{C1one}) can be calculated in a similar way.
Thus, as illustration, we  consider  only one sum:
\begin{equation}
\frac 1\beta \sum_{\omega _m}D_{qz}^{EE}(\mathbf{q},\omega _m)D_{zq}^{EE}(-
\mathbf{q},i\omega _n-i\omega _m)  \label{C1two}
\end{equation}
According to the Lehmann representation, the Green's function can be written in
the form
\begin{equation}
D_{\alpha \beta }^{EE}(\omega _n,\mathbf{r},\mathbf{r}^{\prime })=\frac 1\pi
\int_{-\infty }^\infty dx\frac{\rho _{\alpha \beta }^{EE}(x,\mathbf{r},%
\mathbf{r}^{\prime })}{x-i\omega _n}  \label{Lehmann}
\end{equation}
where
\[
\rho _{\alpha \beta }^{EE}(\omega ,\mathbf{r},\mathbf{r}^{\prime
})=\sum_{n,m}\exp (F-E_n)(E_\alpha (\mathbf{r}))_{nm}(E_\alpha (\mathbf{r}
^{\prime }))_{mn}(1-e^{-\beta\omega _{mn}})\delta (\omega -\omega _{mn})
\]
Using (\ref{Lehmann})  and standard rules for the evaluation of the
sum like (\ref{C1two}) \cite{Mahan} we get
\[
\frac 1\beta \sum_{\omega _m}D_{qz}^{EE}(\mathbf{q},\omega _m)D_{zq}^{EE}(-
\mathbf{q},i\omega _n-i\omega _m)
\]
\[
=\int_{-\infty }^\infty d\omega \Big[\left( \rho _{qz}^{EE}(\mathbf{q},\omega
)D_{zq}^{EE}(-\mathbf{q},i\omega _n-\omega )\right) n(\omega )
\]
\begin{equation}
+\left( D_{qz}^{EE}(
\mathbf{q},i\omega _n-\omega )\rho _{zq}^{EE}(-\mathbf{q},\omega )\right)
(n(\omega )+1)\Big]  \label{C1three}
\end{equation}
Using Eqs.(\ref{b7}) and (\ref{b8}) in (\ref{C1three}) we get
\[
\frac \hbar \beta \sum_{\omega _m}D_{qz}^{EE}(\mathbf{q},\omega
_m)D_{zq}^{EE}(-\mathbf{q},i\omega _n-i\omega _m)=
\]
\[
-q^2\frac \hbar \pi \int_{-\infty }^\infty d\omega \Big[\left( \frac
\partial {\partial z^{\prime }}\frac{\rho _{qq}^{EE}(\omega
,z,z^{\prime })}{\gamma ^2(\omega )}\frac \partial {\partial z}\frac{D_{qq}
(i\omega _n-\omega ,z,z^{\prime })}{\gamma ^2(i\omega
_n-\omega )}\right)n(\omega) 
\]
\begin{equation}
+\left( \frac \partial {\partial z^{\prime }}\frac{D_{qq}^{EE}
(i\omega _n-\omega ,z,z^{\prime })}{\gamma ^2(i\omega _n-\omega )}\frac
\partial {\partial z}\frac{\rho _{qq}^{EE}(\omega ,z,z^{\prime
})}{\gamma ^2(\omega )}\right) (n(\omega )+1)\Big]  \label{C1four}
\end{equation}
where $\gamma^2(\omega )=(\omega /c)^2-q^2$. Replacing 
 $i\omega _n\rightarrow \omega _0+i\delta $ and taking the imaginary
part of Eq.(\ref{C1four}) gives, in the limit $\omega _0\rightarrow 0$, the
following  contribution to the friction coming from Eq.(\ref{C1four})
\[
\lim_{\omega _0\to 0}\mathrm{Im}\frac 1{\omega _0}\lim_{i\omega _n\rightarrow \omega
_0+i\delta }\frac \hbar \beta \sum_{\omega _m}D_{qz}^{EE}(\mathbf{q},\omega
_m)D_{zq}^{EE}(-\mathbf{q},i\omega _n-i\omega _m)=
\]
\begin{equation}
-\frac{2\hbar q^2}{\pi \gamma ^4}\int_{-\infty }^\infty d\omega \left( -
\frac{\partial n}{\partial \omega }\right) \left( \frac \partial {\partial
z}\ \mathrm{Im}D_{qq}(\omega )\right) \left( \frac \partial {\partial
z^{\prime }}\ \mathrm{Im}D_{qq}(\omega )\right)
\end{equation}
Performing a  similar calculation of the other sums in Eq(\ref{C1one}) gives
\[
\gamma _{xx}=\frac \hbar {8\pi ^3}\int_0^\infty d\omega \left( -\frac{
\partial n}{\partial \omega }\right) \int \frac{d^2\mathbf{q}}{(2\pi )^2}
\frac{q_x^2}{q^2}\Big\{\Big[\mathrm{Im}D_{qq}\mathrm{Im}D_{zz}
\]
\[
-\frac{q^2}{\gamma ^4}\left( \frac \partial {\partial z}\mathrm{Im}
D_{qq}\right) \left( \frac \partial {\partial z^{\prime }}\mathrm{Im}
D_{qq}\right) \Big]+
\left( \frac c\omega \right) ^4q^2\Big[\mathrm{Im}D_{nn}\frac{\partial ^2}{
\partial z\partial z^{\prime }}\mathrm{Im}D_{nn}
\]
\begin{equation}
-\left( \frac \partial {\partial z}\mathrm{Im}D_{nn}\right) \left( \frac
\partial {\partial z^{\prime }}\mathrm{Im}D_{nn}\right) \Big]\Big\}%
_{z=z^{\prime }=0}  \label{C1five}
\end{equation}

Using Eqs.(\ref{b7},\ref{b8},\ref{b10},\ref{b11}) for Green's functions
 in Eq.(\ref{C1five}), the contribution to the friction 
from the propagating ($q<\omega/c$) waves becomes: 
\[
\gamma _{\|}^{rad}=\frac \hbar {8\pi ^3}\int_0^\infty d\omega \left( -\frac{%
\partial n}{\partial \omega }\right) \int_{q<\omega /c}d^2\mathbf{q}%
q_x^2\times
\]
\[
\Big[\mathrm{Re}\left( \frac{1+R_{1p}R_{2p}e^{2i\gamma
d}-R_{1p}-R_{2p}e^{2i\gamma d}}{1-e^{2i\gamma d}R_{1p}R_{2p}}\right) \mathrm{%
Re}\left( \frac{1+R_{1p}R_{2p}e^{2i\gamma d}+R_{1p}+R_{2p}e^{2i\gamma d}}{%
1-e^{2i\gamma d}R_{1p}R_{2p}}\right) -
\]
\[
\left( \mathrm{Im}\frac{R_{1p}-R_{2p}e^{2i\gamma d}}{1-e^{2i\gamma
d}R_{1p}R_{2p}}\right) ^2+[p\rightarrow s]\Big]
\]
\begin{equation}
=\frac \hbar {8\pi ^2}\int_0^\infty d\omega \left( -\frac{\partial n}{%
\partial \omega }\right) \int_0^{\omega /c}dqq^3
\frac{(1-|R_{1p}|^2)(1-|R_{2p}|^2)}{|1-e^{2i\gamma d}R_{1p}R_{2p}|^2}%
+[p\rightarrow s]  \label{C1seven}
\end{equation}
Similarly,  the contribution to the friction from the evanescent electromagnetic waves
($q>\omega/c$):
\[
\gamma _{\|}^{evan}=\frac \hbar {8\pi ^3}\int_0^\infty d\omega \left( -\frac{%
\partial n}{\partial \omega }\right) \int_{q<\omega /c}d^2\mathbf{q}%
q_x^2
\]
\[
\times\Big[-\mathrm{Im}\left( \frac{2R_{1p}R_{2p}e^{-2|\gamma
|d}-R_{1p}-R_{2p}e^{-2|\gamma |d}}{1-e^{-2|\gamma |d}R_{1p}R_{2p}}\right)
\mathrm{Im}\left( \frac{2R_{1p}R_{2p}e^{-2|\gamma
|d}+R_{1p}+R_{2p}e^{-2|\gamma |d}}{1-e^{-2|\gamma |d}R_{1p}R_{2p}}\right) 
\]
\[
-\left( \mathrm{Im}\frac{R_{1p}-R_{2p}e^{-2|\gamma |d}}{1-e^{-2|\gamma
|d}R_{1p}R_{2p}}\right) ^2+[p\rightarrow s]\Big]
\]
\begin{equation}
=\frac \hbar {2\pi ^2}\int_0^\infty d\omega \left( -\frac{\partial n}{
\partial \omega }\right) \int_{\omega /c}^\infty dqq^3e^{-2|\gamma |d}
\frac{\mathrm{Im}R_{1p}\mathrm{Im}R_{2p}}{|1-e^{-2|\gamma |}dR_{1p}R_{2p}|^2}
+[p\rightarrow s]
\label{3.1.2}
\end{equation}
Eqs.(\ref{C1seven},\ref{3.1.2}) were first derived in Ref.\cite{Volokitin6} 
using the dynamical 
modification of the Lifshitz theory with use of the fluctuation-dissipation 
theorem.

\subsection{Normal relative motion}

For two plane surfaces in normal relative motion the operator of the force
is given by
\begin{equation}
\hat{F}_z=\int dS_z\hat{\sigma}_{zz},
\end{equation}
where
\begin{equation}
\hat{\sigma}_{zz}=\left(E_zE_z-E_xE_x-E_yE_y+B_zB_z - B_xB_x -B_yB_y\right)/8\pi
\end{equation}

The friction coefficient for normal relative motion can be obtained from the 
analytical continuation of the Green function $G_{zz}(i\omega_m)$ 
which is determined by
\[
G_{zz}(i\omega _n)=\frac{\hbar A}{32\pi ^2\beta }\int \frac{d^2\mathbf{q}}{
(2\pi )^2}\sum_{\omega _m}\frac{q_x^2}{q^2}\Big[D_{zz}^{EE}D_{zz}^{EE}
+D_{qq}^{EE}D_{qq}^{EE} + D_{nn}^{EE}D_{nn}^{EE}
\]
\[
- D_{zq}^{EE}D_{zq}^{EE} - D_{qz}^{EE}D_{qz}^{EE}  - D_{zn}^{EB}D_{zn}^{EB}
-  D_{nz}^{EB}D_{nz}^{EB} 
\]
\begin{equation}
+  D_{qn}^{EB}D_{qn}^{EB} +  D_{nq}^{EB}D_{nq}^{EB} + [E \leftrightarrow 
B]\Big]  \label{Cone}
\end{equation}
where the arguments of the Green functions in Eq.(\ref{Cone}) are the same as in 
Eq.(\ref{C1one}), $[E \leftrightarrow
B]\Big]$ denotes the terms which can be obtained from the first terms by 
permutation of $E$ and $B$ in the upper case indexes of the Green functions, and
\begin{equation}
D_{ij}^{EB}(\mathbf{r}, \mathbf{r}^{\prime}, \omega_n) =
\frac {c}{\omega_n}\mathrm{rot}_{jl}^{\prime}
D_{ij}^{EE}(\mathbf{r}, \mathbf{r}^{\prime}, \omega_n)
\end{equation}
\begin{equation}
D_{ij}^{BE}(\mathbf{r}, \mathbf{r}^{\prime}, \omega_n) =
-\frac {c}{\omega_n}\mathrm{rot}_{il}
D_{lj}^{EE}(\mathbf{r}, \mathbf{r}^{\prime}, \omega_n)
\end{equation} 
Performing similar calculations as  for the parallel relative motion we get
\[
\gamma _{\perp}=\frac \hbar {16\pi ^3}\int_0^\infty d\omega \left( -\frac{
\partial n}{\partial \omega }\right) \int \frac{d^2\mathbf{q}}{(2\pi )^2}
\]
\[
\times\Big\{\Big[\left( \mathrm{Im}D_{qq}\right) ^2+\frac{\gamma ^4}{q^4}\left(
\mathrm{Im}D_{zz}\right) ^2+
\frac 2{\gamma ^2}\left( \frac \partial {\partial z}\mathrm{Im}D_{zz}\right)
^2\Big]
\]
\[
+\left( \frac c\omega \right) ^4\Big[\gamma
^4\left( \mathrm{Im}D_{nn}\right) ^2+\left(
\frac{\partial ^2}{\partial z\partial z^{\prime }}\mathrm{Im}D_{nn}\right)
^2
\]
\begin{equation}
+2\gamma ^2\left( \frac \partial {\partial z^{\prime }}\mathrm{Im}%
D_{nn}(z,z^{\prime })\right) ^2\Big]\Big\}_{z=z^{\prime }=0}  \label{C2one}
\end{equation}
Substitution the  expressions for the Green's functions
from Eqs.(\ref{b7},\ref{b8},\ref{b10},\ref{b11}) in Eq.(\ref{C2one}) gives 
the contribution to the friction from
propagating waves:
\[
\gamma _{\perp}^{rad}=\frac \hbar {16\pi ^3}\int_0^\infty d\omega \left( -\frac{%
\partial n}{\partial \omega }\right) \int_{q<\omega /c}d^2\mathbf{q}\gamma
^2\times
\]
\[
\Big[\left( \mathrm{Re}\frac{1+R_{1p}R_{2p}e^{2i\gamma
d}-R_{1p}-R_{2p}e^{2i\gamma d}}{1-e^{2i\gamma d}R_{1p}R_{2p}}\right)
^2+\left( \mathrm{Re}\frac{1+R_{1p}R_{2p}e^{2i\gamma
d}+R_{1p}+R_{2p}e^{2i\gamma d}}{1-e^{2i\gamma d}R_{1p}R_{2p}}\right) ^2
\]
\[
+2\left( \mathrm{Im}\frac{R_{1p}-R_{2p}e^{2i\gamma d}}{1-e^{2i\gamma
d}R_{1p}R_{2p}}\right) ^2+[p\rightarrow s]\Big]=
\frac \hbar {4\pi ^2}\int_0^\infty d\omega \left( -\frac{\partial n}{
\partial \omega }\right) \int_0^{\omega /c}dqq\gamma ^2
\]
\begin{equation}
\times\frac{(1-|R_{1p}|^2|R_{2p}|^2)^2+|(1-|R_{1p}|^2)R_{2p}e^{i\gamma
d}+(1-|R_{2p}|^2)R_{1p}^{*}e^{-i\gamma d}|^2}{|1-e^{2i\gamma
d}R_{1p}R_{2p}|^4}+[p\rightarrow s]  \label{3.2.1}
\end{equation}
In similar way one can obtain the contribution to the friction from the 
evanescent electromagnetic waves:
\[
\gamma _{\perp}^{evan}=\frac \hbar {4\pi ^3}\int_0^\infty d\omega \left( -\frac{%
\partial n}{\partial \omega }\right) \int_{q>\omega /c}d^2\mathbf{q}|\gamma
|^2
\]
\[
\times\Big[\left( \mathrm{Im}\frac{2R_{1p}R_{2p}e^{-2|\gamma
|d}-R_{1p}-R_{2p}e^{-2|\gamma |d}}{1-e^{-2|\gamma |d}R_{1p}R_{2p}}\right)
^2+\left( \mathrm{Im}\frac{2R_{1p}R_{2p}e^{-2|\gamma
|d}+R_{1p}+R_{2p}e^{-2|\gamma |d}}{1-e^{-2|\gamma |d}R_{1p}R_{2p}}\right)
^2
\]
\[
-2\left( \mathrm{Im}\frac{R_{1p}-R_{2p}e^{-2|\gamma |d}}{1-e^{-2|\gamma
|d}R_{1p}R_{2p}}\right) ^2+[p\rightarrow s]\Big]=
\frac \hbar {\pi ^2}\int_0^\infty d\omega \left( -\frac{\partial n}{\partial
\omega }\right) \int_{\omega /c}^\infty dqq|\gamma |^2e^{-2|\gamma |d}
\]
\[
\times\lbrack (\mathrm{Im}R_{1p}+e^{-2|\gamma |d|}|R_{1p}|^2\mathrm{Im}R_{2p})(
\mathrm{Im}R_{2p}+e^{-2|\gamma |d|}|R_{2p}|^2\mathrm{Im}R_{1p})
\]
\begin{equation}
+e^{-2|\gamma |d}(\mathrm{Im}(R_{1p}R_{2p}))^2]\frac 1{|1-e^{-2|\gamma
|d}R_{1p}R_{2p}|^4}+[p\rightarrow s]
\label{3.2.2}
\end{equation}
Eqs.(\ref{3.2.1},\ref{3.2.2}) were first presented without derivation in 
Ref.\cite{Volokitin4}. The derivation was given in Ref.\cite{Volokitin5} 
on the base of the dynamical modification 
of the semiclassical Lifshitz theory \cite{Lifshitz2} 
of the van der Waals interaction
and the Rytov theory \cite{Rytov,Levin2,Rytov1}
of the fluctuating electromagnetic field.

\section{Van der Waals friction between a small particle and plane surface
obtained using quantum field theory}

\subsection{Parallel relative motion}

Let us assume that the $xy$- plane coincides with surface of the body and 
that the $z$-
axes is directed along the upward normal. For parallel relative motion the
friction coefficient $\Gamma_{\|}=\Gamma_{xx}= \Gamma_{yy}$. The Lorentz
force acting on a small particle located at point $\mathbf{r}_0$ can be
written in the form
\begin{equation}
\hat{F}_x=\left[p_k\frac{\partial}{\partial x_k}E_x(\mathbf{r})+\frac{1}{c}
\left(j_yB_z-j_zB_y\right)\right]_{\mathbf{r}=\mathbf{r}_0}
\label{particle1}
\end{equation}
where $\mathbf{p}$ and $\mathbf{j}$ are  the dipole moment
and current operators of the particle, respectively. 
$\mathbf{E}$ and $\mathbf{B}$ are
 the external electric and magnetic induction field operators,
respectively. The interaction of the electromagnetic field with the particle
is described by the Hamiltonian
\begin{equation}
H_{int}=-\frac{1}{c}\mathbf{A}(\mathbf{r}_0)\cdot \mathbf{j}
\label{particle2}
\end{equation}
where $\mathbf{A}(\mathbf{r})$ is  the vector potential operator.
Taking into account that
\begin{equation}
\mathbf{j}=\frac \partial {\partial t}\mathbf{p}
\end{equation}
\begin{equation}
\mathbf{\nabla }\times \mathbf{E}=\frac 1c\frac \partial {\partial t}\mathbf{%
B}
\end{equation}
one can prove that the friction coefficient is determined by Eq.(\ref{d1}) 
 where
\[
G_{xx}^R(\omega) =
\frac i\hbar \int_0^\infty dte^{i\omega t}
\]
\begin{equation}
\times\Big<p_k(t)\frac \partial
{\partial x}E_k(\mathbf{r},t)p_l(0)\frac \partial {\partial x^{\prime
}}E_l(\mathbf{r}^{\prime },0)-
p_l(0)\frac \partial {\partial x^{\prime }}E_l(\mathbf{r}^{\prime
},0)p_k(t)\frac \partial {\partial x}E_k(\mathbf{r},t)\Big>_{\mathbf{r} =\mathbf{r}
^{\prime}=\mathbf{r}_0}
\label{particle4}
\end{equation}
where  summation over repeated indexes is assumed. 
After similar calculations as in Sec.3 we get
\[
\Gamma_{\|}=
\frac{2\hbar}{\pi}\int_0^{\infty}d\omega\left(-\frac{\partial n}{\partial
\omega}\right) \Big\{\sum_{k=x,y,z}
\mathrm{Im}\alpha_{kk}\frac{\partial^2}{\partial x \partial x^{\prime}}
\mathrm{Im}D_{kk}(\mathbf{r},\mathbf{r}^{\prime},\omega)
\]
\begin{equation}
-2 \mathrm{Re}\left(\alpha_{xx}(\omega)\alpha_{zz}^*(\omega)\right) \left(%
\frac{\partial}{\partial x}\mathrm{Im}D_{xz}(\mathbf{r},\mathbf{r}
_0,\omega) \right)^2\Big\}_{\mathbf{r}=\mathbf{r}^{\prime}=\mathbf{r}_0}
\label{particle8}
\end{equation}
where $D_{ij}(\mathbf{r},
\mathbf{r}^{\prime})$ are the Green's functions of the electromagnetic field 
for one plane surface  geometry  without interaction with a particle. 
These Green functions can be obtained from the Green functions for two 
plane surface geometry (see Appendix A) by putting $R_{p(s)}=0$.   
The polarizability of the particle 
\begin{equation}
\alpha_{kk}(\omega)=\frac i\hbar \int_0^\infty dte^{i\omega t}
<p_k(t)p_k(0)-p_k(0)p_k(t)>
\end{equation}
is expressed through its value  $\alpha_{kk}^0$ in the absence
of interaction between the  particle and the surface
\begin{equation}
\alpha_{ii}(\omega)=\frac{\alpha_{ii}^0(\omega)}{1-\alpha_{ii}^0(\omega)
D_{ii}(\mathbf{r}_0,\mathbf{r}_0)}.
\end{equation}
We have also used the identity
\[
\mathrm{Im}\alpha_{xx}(\omega)\mathrm{Im}\left[\alpha_{zz}(\omega) \frac{%
\partial}{\partial x}D_{xz}(\mathbf{r},\mathbf{r}_0,\omega) \frac{\partial%
}{\partial x}D_{xz}(\mathbf{r} ,\mathbf{r}_0,\omega)\right]
\]
\[
+\mathrm{Im}\alpha_{zz}(\omega)\mathrm{Im}\left[\alpha_{xx}(\omega) \frac{
\partial}{\partial x}D_{xz}(\mathbf{r},\mathbf{r}_0,\omega) \frac{\partial
}{\partial x}D_{xz}(\mathbf{r} ,\mathbf{r}_0,\omega)\right]
\]
\[
-2\mathrm{Im}\left[\alpha_{xx}(\omega) \frac{\partial}{\partial x}D_{xz}(
\mathbf{r},\mathbf{r}_0,\omega)\right] \mathrm{Im}\left[\alpha_{zz}(\omega)
\frac{\partial}{\partial x}D_{xz}(\mathbf{r},\mathbf{r}_0,\omega)\right]
\]
\begin{equation}
=2\mathrm{Re}\left(\alpha_{xx}(\omega)\alpha_{zz}(\omega)^*\right) \left(
\frac{\partial}{\partial x}\mathrm{Im}D_{xz}(\mathbf{r},\mathbf{r}%
_0,\omega) \right)^2  
\end{equation}

\subsection{Normal relative motion}

The friction coefficient for a particle moving  normal to the sample surface can be 
obtained from calculations very similar to those for parallel relative  motion. 
In this case the Green's function $G_{xx}^R$ must be replaced by $G_{zz}^R$ where
\[
G_{zz}^R(\omega) =
\frac i\hbar \int_0^\infty dte^{i\omega t}
\]
\begin{equation}
\times\Big<p_k(t)\frac \partial
{\partial z}E_k(\mathbf{r},t)p_l(0)\frac \partial {\partial z^{\prime
}}E_l(\mathbf{r}^{\prime },0)-
p_l(0)\frac \partial {\partial z^{\prime }}E_l(\mathbf{r}^{\prime
},0)p_k(t)\frac \partial {\partial z}E_k(\mathbf{r},t)\Big>_{\mathbf{r} =\mathbf{r}
^{\prime}=\mathbf{r}_0}
\label{e2}
\end{equation}
After similar calculations as in Sec.4.1 we get
\[    
\Gamma_{\perp}=
\frac{2\hbar}{\pi}\int_0^{\infty}d\omega\left(-\frac{\partial n}{\partial
\omega}\right) \sum_{k=x,y,z}\Big\{
\mathrm{Im}\alpha_{kk}(\omega)\frac{\partial^2}{\partial z \partial
z^{\prime}} \Big[\mathrm{Im}D_{kk}(\mathbf{r},\mathbf{r}^{\prime},\omega)
\]
\begin{equation}
+\mathrm{Im}\left(\alpha_{kk}D_{kk}(\mathbf{r},\mathbf{r}_0,\omega)
D_{kk}(\mathbf{r}^{\prime},\mathbf{r}_0,\omega)\right)\Big]+ \Big[\frac{%
\partial}{\partial z}\mathrm{Im}\left(\alpha_{kk}(\omega) D_{kk}(\mathbf{r}%
,\mathbf{r}_0,\omega)\right)\Big]^2\Big\}_{\mathbf{r}=\mathbf{r}^{\prime}=%
\mathbf{r}_0}  \label{particle9}
\end{equation}
For a spherical particle with radius $R$ Eq.(\ref{particle9}) is only valid if 
$R\ll d$. In non-resonant case $\alpha_{kk}^0 \sim R^3$ and $D_{kk} \sim d^{-3}$.
Thus in this case we can neglect by the screening effects. Than taking into account 
that for a spherical particle $\alpha_{kk}=\alpha$, and using in the non-retarded limit 
(which can be formally obtained as a limit $c \rightarrow \infty$) formula (see 
\cite{Volokitin7} and also Appendix A)
\begin{equation}
\sum_{k=x,y,z}D_{kk}(\mathbf{r}, \mathbf{r}^{\prime}, \omega) = 
4\pi\int\frac{d^2qq}{(2\pi)^2}\left[e^{-q|z-z^{\prime}|}+
R_p(q,\omega)e^{-q(z+z^{\prime})}\right]e^{i\mathbf{q}(\mathbf{x}-
\mathbf{x}^{\prime})}
\end{equation}
we get 
\begin{equation}
\Gamma _{\| }=2\frac {\hbar}{ \pi} \int_0^\infty \mathrm{d}\omega \left( -
\frac{\partial n(\omega )}{\partial \omega }\right) \int_0^\infty \mathrm{d}
qq^4e^{-2qd}
\mathrm{Im}R_p(q,\omega )\mathrm{Im}\alpha (\omega )  \label{sparticle2}
\end{equation}
and $\Gamma_{\perp}=2\Gamma_{\|}$. Eqs.(\ref{particle8},{particle9}) 
were first obtained 
in Ref.\cite{Volokitin7} with use of the dynamical modification of the semiclassical 
Rytov theory \cite{Rytov,Levin2,Rytov1} of the fluctuating electromagnetic field. 
The particular case of these equations given by Eq.(\ref{sparticle2}) was first 
derived in Ref.\cite{Tomassone} using the fluctuation-dissipation theorem.

\section{Van der Waals friction between dielectrics and two-dimensional systems}

In Refs.\cite{Persson and Volokitin,Volokitin5} it was shown 
that the van der Waals friction between 
good conductors ($k_BT/4\pi \sigma \gg 1$, where $\sigma$ is the conductivity)
 is extremely small. 
However the van der Waals friction can be greatly enhanced for high 
resistivity materials.
Recently \cite{Greffet2} the calculations were published, where the authors
claim that they can explain the experimental data about long-range
noncontact friction observed by Stipe \textit{et.al} \cite{Stipe}. In this
experiment the substrate did not consist of a bulk conductor but of a metal
film with a thickness of a few hundred nanometers deposited on top of a
dielectric substrate. Thus in the theory from Ref.\cite{Greffet2} it was
proposed that experimental measurements performed on metal films do not
reflect the properties of the metal but of the underlying dielectric
substrate. Using the formula (\ref{sparticle2})  with the particle polarazability 
\begin{equation}
\alpha(\omega) = R^3 \frac{\varepsilon - 1}{\varepsilon + 2}
\end{equation}
and the reflection coefficient in the electrostatic limit ($q\gg \omega/c$)
\begin{equation}
R_p= \frac{\varepsilon - 1}{\varepsilon + 1},
\end{equation}
 for
high- resistivity material ($4\pi \sigma<<k_{B}T/\hbar$) we get: 
\begin{equation}
\Gamma_{\|} =\frac{9}{2}\frac{k_BT}{4\pi\sigma}\frac{R^3}{d^5} \frac{1}{
2\varepsilon^{\prime} +3}  \label{greffet2}
\end{equation}
where $\varepsilon^{\prime}$ is the real part of the dielectric function $
\varepsilon=\varepsilon^{\prime}+4\pi i\sigma/\omega$. In the frequency
range below $0.75\times10^{12}$Hz the real part of the dielectric constant
of glass is nearly constant ($\varepsilon^{\prime} \approx 3.82$), and $%
\sigma=30$s$^{-1}$. It was assumed that $d=s+h_s+h_t$, where $s$ is the
tip's apex-surface separation, and $h_s$ and $h_t$ are the thicknesses of
the gold films which were deposited on the mica substrate and on the silicon
cantilever, respectively. In experiment \cite{Stipe} $h_s=250$nm, $h_t=200$nm, 
 $
R=1\mu$m, than for $s=10$nm and $T=300$K we get 
$\Gamma_{\|} = 7\times 10^{-13}$kg/s, which agrees exactly with value
obtained numerically in \cite{Greffet2}, and in rough agreement with the
experimental value reported in Ref.\cite{Stipe}. However
 the macroscopic theory which was used in
obtaining Eq.(\ref{greffet2}) is only valid when the average separation 
between electrons is much smaller than length scale of variation of the 
electric field, which is determined by separation $d$. Thus the lowest 
value of the electron concentration $n_{min}$ is restricted by the condition 
$n_{min}\geq d^{-3}$ and according to Drude formula the lowest value 
of the conductivity  $\sigma \geq
\sigma_{min}\sim e^2\tau/md^3$. For $d\sim 1\mu$m and $\tau=10^{-15}$ the
conductivity must be at least  four orders of magnitude larger than it was
used in \cite{Greffet2}. For such conductivity Eq.(\ref{greffet2}) gives $%
\Gamma_{\|} \sim 10^{-16}$kg/s, which is three orders of the magnitude smaller
than the observed friction. 

Besides, in the experiment \cite{Stipe} the mica substrate and silica tip
were coated by gold films thick enough to completely screen the
electrodynamic interaction between underlying dielectrics. If planar film
with thickness $h$ and dielectric function $\varepsilon_3(\omega)$ lies on a
substrate with dielectric function $\varepsilon_2(\omega)$, the system
response is similar to the single-interface case if one replace the
reflection coefficients $R_{p21}$ by 
\begin{equation}
R_p=\frac{R_{p31}-R_{p32}\exp(-2qh)}{1-R_{p31}R_{p32}\exp(-2qh)}
\label{coefficient}
\end{equation}
where 
\begin{equation}
R_{pij}=\frac{\varepsilon_j-\varepsilon_i}{\varepsilon_j+
\varepsilon_i}
\end{equation}. The magnitude of the wave vector $q$ is determined by the 
separation between metallic films. Thus $qh  \sim h/s \gg 1$, and in 
this case reflection 
coefficient $R_p \approx R_{31}$ instead of $R_{21}$ as it was proposed in 
Ref.\cite{Greffet2}.

For high-resistivity metals ($k_BT/4\pi \hbar \sigma >1$), 
for $d<c(\hbar/4\pi\sigma/ll 1$ 
the reflection coefficient (\ref{5.1}) in Eq.(\ref{3.1.2}) 
 for two surfaces in parallel relative we get 
\begin{equation}
\gamma _{\|}\approx 0.05\frac \hbar {d^4}\frac{k_BT}{4\pi \hbar \sigma }
\end{equation}
and $\gamma _{\perp }\approx 10 \gamma _{\| }$. 
From the condition of the validity of the macroscopic theory (see above) 
 maximum of friction can be estimated as
\begin{equation}
\gamma _{\|max}\sim 0.1 \frac \hbar {d^4}\frac{k_BT}{4\pi \hbar \sigma _{min}}\sim
\frac{mk_BT}{4\pi e^2\tau d}.
\end{equation}

To estimate the friction coefficient $\Gamma $ for an atomic force
microscope tip with the radius of curvature $R>>d$ we can use the ``proximity 
 approximation'' \cite{Hartmann,Apell}.
 Thus the maximal friction
coefficient for a spherical tip:
\begin{equation}
\Gamma _{\|max}^s\sim 0.1 \gamma _{max}dR\sim \frac{mk_BTR}{4\pi e^2\tau }.
\end{equation}
For $\tau \sim 10^{-16}$s, $R\sim 1\mu$m and $T=300$K we get $\Gamma
_{max}\sim 10^{-15}$kg/s. This friction is two order of magnitude smaller
than the friction observed in a recent experiment at $d=10$nm \cite{Stipe}.
Similarly, in the case of a cylindrical tip:
\begin{equation}
\Gamma _{max}^c\sim \gamma _{max}\sqrt{dR}w\sim \frac{mk_BTR^{1/2}w}{4\pi
e^2\tau d^{1/2}}
\end{equation}
where $w$ is the width of the tip. For $w=7\mu $m, $d=10$nm, and with the
other parameters as above, the friction is  of the same orders of magnitude as it was 
observed in the experiment \cite{Stipe}.

Recently a large
electrostatic non-contact friction between an atomic force microscope tip and
thin dielectric films was observed \cite{Kuehn}.
Similarly the  van der Waals 
friction will be also large between  dielectrics with high absorption 
in  low-frequency range.  As a particular important case we consider 
the van der Waals friction between thin water films adsorbed on the surfaces of the 
transparent dielectric substrates like silica or mica. Water has an extremely large 
static dielectric function of around 80. The low frequency contribution to the 
dielectric function, responsible for this large static value, is due to relaxation 
of the permanent dipoles of the water molecules. It is very nicely described by the 
simple Debye \cite{Debye} rotation relaxation. The theoretical fit of the experimental 
data is given by \cite{Sernelius}:
\begin{eqnarray}
\varepsilon^{\prime}(\omega) = 4.35+\frac {C}{\left(1+(\omega/\omega_0)^2\right)}  \\
\varepsilon^{\prime \prime}(\omega) = \frac {C(\omega/\omega_0)^2}
{\left(1+(\omega/\omega_0)^2\right)}       
\end{eqnarray}
where $C=72.24$ and $\omega_0=1.3\cdot 10^{11}$s$^{-1}$. The fit is very good for both 
the real  and imaginary  
part of the dielectric function $\varepsilon (\omega)=\varepsilon^{\prime}(\omega) +
i\varepsilon^{\prime \prime}(\omega)$ 
We note that water has large absorption in the radio-frequency range at $\omega 
\sim \omega_0$, and shows in this region of the spectrum anomalous dispersion. In this 
frequency range the dielectric constants $\varepsilon_2$ of mica or silica are nearly 
uniform and $|\varepsilon_3| \gg \varepsilon_2$, where $\varepsilon_3$ denotes the 
dielectric function of water. The reflection coefficient for substrate with film is 
given by Eq.(\ref{coefficient}). For $qh \ll 1$ and $q^{-1}\sim d \ll |\varepsilon_3|h/
\varepsilon_2$ it can be approximated by equation
\begin{equation}
R_p \approx 1-\frac{2}{\varepsilon_3qh}
\label{4.1}
\end{equation}   
Substituting (\ref{4.1}) in eq.(\ref{3.1.2}) and using ``proximity''
 approximation for 
friction between cylindrical atomic force microscope tip and a sample we get
\begin{equation}
\Gamma_{\|}^c = \frac{\pi\hbar R^{1/2}w}{6\sqrt{2}C^2h^2d^{3/2}}
\left(\frac{k_BT}{\hbar \omega_0}\right)^2
\label{4.2}
\end{equation}
For $h=1$nm and with the same parameters as above we get $\Gamma_{\|}^c=4.8\cdot
10^{-12}$kg/s what is one order of magnitude larger than the friction observed in 
Ref.\cite{Stipe}. It is interesting to note that the friction coefficient 
(\ref{4.2}) has the same weak distance dependence as it 
was observed in Ref.\cite{Stipe}.

Another enhancement mechanism of the van der Waals friction is connected
with resonant photon tunneling between adsorbate vibrational modes localized
on different surfaces. In \cite{Volokitin4,Volokitin5} we have shown that
resonant photon tunneling between two surfaces separated by $d=1$nm, and
covered by a low concentration of potassium atoms, result in 
a friction which is six orders of the magnitude larger than for clean surfaces. The
adsorbate induced enhancement of the van der Waals friction is even larger
for Cs adsorption on Cu(100). In this case, even at  low coverage ($
\theta \sim 0.1$), the adsorbed layer exhibits an acoustic branch for
vibrations parallel to the surface \cite{Senet}. Thus, $\omega _{\Vert }=0$
and according to Ref.\cite{Volokitin9} at
small frequencies the reflection coefficient is given by
\begin{equation}
R_p=1-\frac{2qa\omega _q^2}{\omega ^2-\omega _q^2+i\omega \eta }
\label{adsorbate1}
\end{equation}
where $\omega _q^2=4\pi n_ae^{*2}aq^2/M$, $e^{*}$ is the ion charge and $a$
is the separation between an ion and the image plane. Substituting Eq.(\ref
{adsorbate1}) in Eq. (\ref{3.1.2}) for
\[
\frac a{\eta d}\sqrt{\frac{4\pi n_ae^{*2}a}{Md^2}}\ll 1,
\]
and using a ``proximity'' approximation, for a cylindrical tip we get
\begin{equation}
\Gamma _{\Vert }^c\approx 0.68\frac{k_BTa^2R^{0.5}w}{\eta d^{5.5}}
\label{adsorbate3}
\end{equation}
For Cs adsorption on Cu(100) the damping parameter $\eta $ was estimated in
\cite{Volokitin9} as $\eta \approx  10^{11}$s$^{-1}$. Using this value of $
\eta $ in Eq.(\ref{adsorbate3}) for $a=2.94${\AA } \cite{Senet}, $R=1\mu $m,
$w=7\mu $m, $T=293$ K at $d=10$nm we get $\Gamma _{\| }\approx\cdot 10^{-15}$
kg/s, which is two orders of magnitude smaller than the friction observed in \cite
{Stipe} at the same distance. However, the van der Waals friction is
characterized by a much stronger distance dependence ($\sim 1/d^{5.5}$) than
observed in the experiment ($\sim 1/d^n$, where $n=1.3\pm 0.2$). Thus, at
small distances the van der Waals friction will be much larger than
friction observed in \cite{Stipe}, and can thus be measured experimentally. 

\section{Summary and conclusion}

We have used a quantum field theory in Matsubara formulation to calculate 
the van der Waals friction between two plane surfaces, and a small particle and plane 
surface, for parallel and normal relative motion.   
The friction calculated in this approach is
in agreement with the friction calculated in the framework of dynamical modification
of the Lifshitz theory with use of the fluctuation-dissipation theorem. This
solves the problem of the applicability of the Lifshitz theory to the dynamical
situation. In quantum field theory the calculation of the friction to linear
order in the sliding velocity is reduced to the finding of the equilibrium
Green functions which obey to the system of the Maxwell type differential equations. 
Thus with application of the numerical methods of classical electrodynamics 
this approach can be used in the calculations of the van der Waals friction
between bodies with complex geometry.
We show that the van der Waals friction between high-resistivity metals, dielectrics 
with strong absorption in radio-frequency range, and two-dimensional systems 
can be measured in non-contact friction
experiment using state-of-the-art equipment. The theory can be used as a guide for 
planning and interpretation of  new experiments including more sophisticated 
dielectrics 
like in Ref.\cite{Kuehn}.

\vskip 0.5cm \textbf{Acknowledgment }

A.I.V acknowledges financial support from  DFG and ESF ``Nanotribology''.

\appendix

\section{The Green functions for two plane surface geometry}

Suppose that the half-space $z<0$ is occupied by a solid 
with the reflection coefficients $R_{1p}(\mathbf{q},\omega )$ and $R_{1s}(
\mathbf{q},\omega )$ for $s$- and $p$- electromagnetic fields, respectively.
Similarly, the half-space $z>d$ is occupied with a solid 
with the reflection coefficients $R_{2p}(\mathbf{q},\omega )$ and $R_{2s}
\mathbf{q},\omega )$. The region $0<z<d$ is assumed to be vacuum.
Here $\mathbf{q}$ is  the surfaces component of wave vector $\mathbf{k}=
(\mathbf{q},\gamma)$, where $\gamma = ((\omega/c)^2-q^2)^{1/2}$.
Since the system is uniform in the $\mathbf{x}=(x,y)$ directions, the Green
function $D_{ij}(\mathbf{r},\mathbf{r}^{{\prime }})$ can be represented by
the Fourier integral
\begin{equation}
D_{ij}(\mathbf{r},\mathbf{r}^{\prime },omega_n)=
\int \frac{\mathrm{d}^2q}{({2\pi )}^2%
}e^{\mathrm{i}\mathbf{q}(\mathbf{x}-\mathbf{x}^{\prime })}D_{ij}(z,z^{\prime
},\mathbf{q},i\omega_n )
\end{equation}

In the $xy-$plane it is convenient to choose the coordinate axes along the
vectors $\hat q=\mathbf{q}/q$ and $\hat n=\hat z\times \hat q$. In this
coordinate system the equations (\ref{green1}, \ref{green2}) for the Green
functions become
\begin{eqnarray}
\left( \gamma^2+\frac{\partial ^2}{\partial z^2}\right) D_{nn}(z,z^{\prime
}) &=&-\frac{4\pi \omega ^2}{c^2}\delta (z-z^{\prime })  \label{b1} \\
\left( \frac{\omega ^2}{c^2}-\frac{\partial ^2}{\partial z^2}\right)
D_{qq}(z,z^{\prime })-\mathrm{i}q\frac \partial {\partial
z}D_{zq}(z,z^{\prime }) &=&-\frac{4\pi \omega ^2}{c^2}\delta (z-z^{\prime })
\label{b2} \\
\gamma ^2D_{zq}(z,z^{\prime })-\mathrm{i}q\frac \partial {\partial
z}D_{qq}(z,z^{\prime }) &=&0  \label{b3} \\
\gamma ^2D_{zz}(z,z^{\prime })-\mathrm{i}q\frac \partial {\partial
z}D_{qz}(z,z^{\prime }) &=&-\frac{4\pi \omega ^2}{c^2}\delta (z-z^{\prime })
\label{b4} \\
\gamma^2D_{qz}(z,z^{\prime })+\mathrm{i}q\frac \partial {\partial z^{\prime
}}D_{qq}(z,z^{\prime }) &=&0  \label{b5}
\end{eqnarray}
The components $D_{qn}$, $D_{zn}$ of the Green function vanish, since the
equations for them turn out to be homogeneous.
Solving the system of Eqs.(\ref{b1})-(\ref{b5}) amounts to solving 
 two equations: equation 
(\ref{b1}) for $D_{nn}$ and the equation for $D_{qq}$ which follows from
equations (\ref{b2}, \ref{b3})
\begin{equation}
\left( \gamma ^2+\frac{\partial ^2}{\partial z^2}\right) D_{qq}(z,z^{\prime
})=-4\pi \gamma^2\delta (z-z^{\prime })  \label{b6},
\end{equation}
after which $D_{qz}$, $D_{zq}$ and $D_{zz}$ for $z\not =z^{\prime }$ are
obtained as
\begin{eqnarray}
D_{qz}^R &=&-\frac{iq}{\gamma ^2}\frac \partial {\partial z^{\prime
}}D_{qq};D_{zq}=\frac{iq}{\gamma ^2}\frac \partial {\partial
z}D_{qq};  \label{b7} \\
D_{zz} &=&\frac{q^2}{\gamma ^4}\frac{\partial ^2}{\partial z\partial
z^{\prime }}D_{qq}  \label{b8}
\end{eqnarray}
In the vacuum gap $0<z<d$ the solution of equation (\ref{b1}) has
the form
\begin{equation}
D_{nn}(z,z^{\prime })=-\frac{2\pi i \omega ^2}{\gamma c^2}e^{i\gamma
\mid z-z^{\prime }\mid }+v_ne^{i\gamma z}+w_ne^{-i\gamma z}
\label{b9}
\end{equation}
At the boundaries  $z=0$ and $z=d$ the amplitude of the scattered wave is
equal to amplitude of incident wave times to corresponding reflection
coefficient. The Green function $D_{nn}$ is associated with $s$-
polarized electromagnetic field and the boundary
conditions for it give
\begin{eqnarray}
v_n &=&R_{1s}\left( w_n+\frac{2\pi i\omega ^2}{\gamma c^2}
e^{i\gamma z^{\prime }}\right) \,\,\,\,\,\,\,\,\,\,\,\,\,\,\,\mbox{for}\qquad z=0 \\
w_n &=&R_{2s}e^{2i\gamma d}\left( v_n+\frac{2\pi i\omega ^2}{\gamma
c^2}e^{-i\gamma z^{\prime }}\right) \qquad \mbox{for}\qquad z=d
\label{b15}
\end{eqnarray}
Using Eqs.(\ref{b9}--\ref{b15}) we get
\begin{eqnarray}
D_{nn}(z,z^{\prime }) &=&-\frac{2\pi i\omega ^2}{\gamma c^2}
\Bigg\{e^{i\gamma \mid z-z^{\prime }\mid }+\frac{R_{1s}R_{2s}e^{2i\gamma
d}\left( e^{i\gamma (z-z^{\prime })}+e^{-i\gamma
(z-z^{\prime })}\right) }{\Delta _s}  \nonumber \\
&&+\frac{R_{1s}e^{i\gamma(z+z^{\prime })}+R_{2s}e^{2i\gamma
 d}e^{-i\gamma (z+z^{\prime })}}{\Delta _s}\Bigg\}
\label{b10} \\
\Delta _s &=&1-e^{2i\gamma d}R_{2s}R_{1s}
\end{eqnarray}
Equation (\ref{b6}) for $D_{qq}$ is similar to equation (\ref{b1}) for $
D_{nn}$, and the expression for $D_{qq}^R$ 
can be obtained from expression (\ref{b10})
 just by replacements of reflection coefficient
\begin{equation}
D_{qq}=\left( \frac{\gamma c}{\omega} \right) ^2D_{nn}\left[ R_s\rightarrow
-R_p\right]  \label{b11}
\end{equation}
In our approach the calculation of reflection coefficient
for $s$- and $p$- polarized waves constitutes  separate problem, which can be
solved taking into account non-local effects. For the local optic case the
reflection coefficients are determined by the well known Fresnel formulas
\begin{equation}
R_{ip}=\frac{\varepsilon _i\gamma -\gamma _i}{\varepsilon _i\gamma +\gamma _i%
},\,\,\;\;\;\;\;R_{is}=\frac{\gamma -\gamma _i}{\gamma +\gamma _i},
\label{b12}
\end{equation}
where $\varepsilon _i\,$ is the complex dielectric constant for body $i$:
\begin{equation}
\gamma _i=\sqrt{\frac{\omega ^2}{c^2}\varepsilon _i-q^2}  \label{b13}
\end{equation}

FIGURE CAPTION

Fig. 1. The electromagnetic waves emitted in the opposite direction by the body
at the bottom will experience  opposite Doppler shift in the reference frame
in which the body at the top is at rest. Due to the frequency dispersion of
the reflection coefficient these electromagnetic waves will reflect differently
from the surface of the body at the top, which give rises to
momentum transfer between the bodies. This momentum transfer is the
origin of the van der Waals friction.

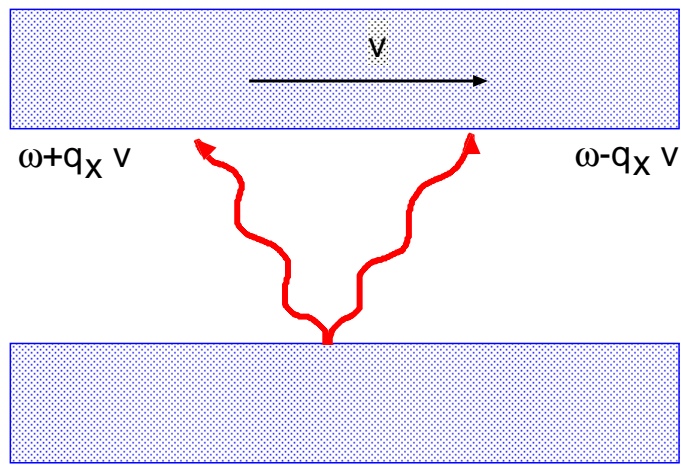
\end{document}